\documentclass[12pt]{article}

\usepackage[english]{babel}
\usepackage{amsmath}
\usepackage{amssymb}
\usepackage{amsthm,amsfonts}
\usepackage{dsfont}
\usepackage{tensor}
\usepackage[dvipsnames]{xcolor}
\usepackage[mathstyleoff]{breqn}
\usepackage{slashed}
\usepackage{stmaryrd}
\usepackage{enumerate}
\usepackage[normalem]{ulem}
\usepackage{tikz}
\usepackage[mathscr]{euscript}
\usepackage{mathrsfs}
\usepackage[numbers,sort&compress]{natbib}
\usepackage{hyperref}
\usepackage{amsthm}
\usepackage{svrsymbols}
\usepackage{phaistos}
\usepackage{mathtools}
\usepackage{stmaryrd} 
\usepackage{graphicx}

\def\be{\begin{equation}}
\def\ee{\end{equation}}
\def\bea{\begin{eqnarray}}
\def\eea{\end{eqnarray}}

\hypersetup{
    colorlinks=true,
    linkcolor=MidnightBlue,
    citecolor=MidnightBlue,
    urlcolor=MidnightBlue
}

\numberwithin{equation}{section} % numbering equations with section number

\begin{document}
	
\pagestyle{plain}

%------------------------------------------------------%
%  title page
%------------------------------------------------------%

\pagestyle{empty}
\begin{center}

\vskip .5cm
		
\LARGE{\bf Braneworld tidal charge through the classical double copy}

\vskip 0.3cm

\normalsize{Juan C. La Cruz$^{a}$ and Jesús A. Rodríguez$^{a,b}$
\\[6mm]}

{$^{a}$ 
\small Universidad Argentina de la Empresa (UADE), Instituto de Tecnología (INTEC),\\ [.01 cm]}
{\small\it Lima 717, Buenos Aires, Argentina\\ [.1 cm]}
{$^{b}$ \small Instituto de Astronomía y Física del Espacio, (CONICET-UBA)\\ [.01 cm]}
{\small\it Ciudad Universitaria, Pabellón IAFE, CABA, C1428ZAA, Argentina\\ [.3 cm]}

{\small \verb"julacruz@uade.edu.ar"\ \ \ \ \  \verb"jarodriguez@iafe.uba.ar"}\\[1.5cm]

\small{\bf Abstract} \\[0.5cm]
\end{center}

%\begin{abstract}
We investigate the classical double copy of the rotating black-hole geometry induced on the brane in the Randall--Sundrum II scenario. The influence of the five-dimensional bulk is encoded in the effective four-dimensional geometry through a tidal charge, while the metric admits a Kerr--Schild representation over a flat background. We construct the corresponding single and zeroth copies and find that their regular sources are entirely proportional to the tidal charge, whereas the mass-dependent contribution is source-free in the regular region. Comparison with Kerr--Newman reveals a formal correspondence between the tidal charge and the squared electromagnetic charge at the level of the induced geometry and its double copies, despite their different physical origins and the fact that the tidal charge may take either sign. When an electromagnetic field is included on the brane, both contributions enter the low-energy Kerr--Schild geometry and its double copies only through the combination $Q^2+\beta$. These results illustrate how physically distinct brane and bulk contributions can become indistinguishable in the classical double copy of an effective geometry.
%\end{abstract}

\newpage
%----------------------------------------------------------------------%
%  Resetting of counters 

%----------------------------------------------------------------------%
\setcounter{page}{1}
\pagestyle{plain}
\renewcommand{\thefootnote}{\arabic{footnote}}
\setcounter{footnote}{0}

% \tableofcontents

% \newpage

%%%%%%%%%%%%%%%%%%%%%%%%%%%%
%%%%%%%% SECTION 1 %%%%%%%%%
%%%%%%%%%%%%%%%%%%%%%%%%%%%%

\section{Introduction}
\label{sec:introduction}

The Randall--Sundrum models provide a realization of extra-dimensional gravity in which the observed hierarchy of physical scales can emerge from a warped five-dimensional spacetime \cite{Randall:1999ee,Randall:1999vf}. In particular, in the Randall--Sundrum II (RSII) scenario \cite{Randall:1999vf}, four-dimensional spacetime is described by a positive-tension 3-brane embedded in an $AdS_5$ bulk with a noncompact extra dimension. The warped geometry localizes gravity near the brane and reproduces four-dimensional Einstein gravity at sufficiently large distances \cite{Randall:1999vf,Garriga:1999yh}.\footnote{For a comprehensive review of braneworld gravity, see \cite{Maartens:2010ar}.}

Black holes localized on the brane require a nontrivial extension of the geometry into the five-dimensional bulk, a problem that arises already for the Schwarzschild-like static black hole and becomes more involved when rotation is included. While fully backreacted static and rotating solutions have been obtained numerically \cite{Figueras:2011gd,Biggs:2021iqw}, analytic constructions require additional assumptions on the bulk matter content \cite{Nakas:2020sey,Nakas:2021srr,Neves:2021dqx,Estrada:2024lhk,Estrada:2025ice}. A closed-form five-dimensional vacuum solution for a rotating black hole localized on the brane, however, remains unknown.

An alternative to constructing the complete five-dimensional geometry is to consider the effective gravitational dynamics induced on the brane. Projecting the five-dimensional Einstein equations onto the brane yields effective four-dimensional equations in which the influence of the bulk is encoded, in part, in the electric component of the five-dimensional Weyl tensor, $E_{\mu\nu}$ \cite{Shiromizu:1999wj}. In the absence of brane matter and for a vanishing effective cosmological constant, the effective geometry is sourced by $E_{\mu\nu}$, whose tracelessness implies $R=0$. Moreover, $E_{\mu\nu}$ carries nonlocal information about the bulk and is not determined by the effective equations alone, allowing black-hole geometries to be constructed on the brane without explicit knowledge of the five-dimensional solution.

Within this effective description, a Reissner--Nordstr\"om-like geometry was obtained in \cite{Dadhich:2000am}, with the squared electric charge replaced by a tidal charge originating from gravitational effects in the bulk. Unlike an ordinary electromagnetic charge, the tidal charge is not constrained to be positive and can therefore lead to gravitational properties with no direct Reissner--Nordstr\"om counterpart. This construction was subsequently extended to the rotating case \cite{Aliev:2005bi}, yielding an induced geometry of the Kerr--Newman form with $Q^2$ replaced by the tidal charge associated with $E_{\mu\nu}$. Of particular interest here, the rotating solution admits a Kerr--Schild representation over a flat background.

The Kerr--Schild structure provides a natural setting for the classical double copy, which relates gravitational solutions to gauge-theory configurations and is motivated by the color--kinematics duality of scattering amplitudes \cite{Bern:2008qj,Bern:2010ue}. For gravitational solutions in Kerr--Schild form, the simplified gravitational equations can be directly related to Maxwell-type equations for a single-copy gauge field constructed from the Kerr--Schild scalar and null vector \cite{Monteiro:2014cda}. Since its original formulation, the Kerr--Schild double copy has been explored in a variety of gravitational settings, including black holes, curved backgrounds, higher-dimensional spacetimes, and extensions involving additional fields and duality-covariant formulations \cite{Luna:2015paa,Ridgway:2015fdl,Bahjat-Abbas:2017htu,
Carrillo-Gonzalez:2017iyj,Lee:2018gxc,Bah:2019sda,Lescano:2020nve,Lescano:2021ooe,Alkac:2021bav,Lescano:2022nhp,Easson:2022zoh,Chawla:2023bsu,Zhao:2024wtn,Albertini:2025ogf,Alencar:2026zdz,Morieri:2026gdo}. For recent reviews of the double copy and its classical realizations, see \cite{Lescano:2026xlb,CarrilloGonzalez:2026phk}.

The interplay between the classical double copy and the RSII scenario was recently explored for the Randall--Sundrum black string \cite{Chamblin:1999by,Rodriguez:2026zxm}, whose five-dimensional geometry is explicitly given by a warped extension of the four-dimensional Schwarzschild solution along the extra dimension. Here we consider instead a rotating black hole localized on the brane, for which the complete bulk geometry is not known analytically and its influence is encoded in the effective four-dimensional geometry through $E_{\mu\nu}$ \cite{Aliev:2005bi}. The analogy between the tidal charge $\beta$ and the Kerr--Newman charge parameter $Q^2$ then becomes particularly relevant in the context of the classical double copy, raising the question of whether their different physical origins can be distinguished by the corresponding single and zeroth copies.

In this work, we analyze the Kerr--Schild structure of the rotating braneworld black hole and construct its single and zeroth copies, showing that their regular sources are entirely controlled by the tidal charge. Comparison with Kerr--Newman reveals a formal correspondence $Q^2\leftrightarrow\beta$ between the two double copies, despite the different physical origins of the electromagnetic and tidal charges. We then introduce an electromagnetic field on the brane and show that, in the low-energy regime, both contributions enter the induced Kerr--Schild geometry and its double copies only through the combination $Q^2+\beta$. Beyond this approximation, the quadratic contribution of the brane energy-momentum tensor modifies this simple degeneracy.

%%%%%%%%%%%%%%%%%%%%%%%%%%%%
%%%%%%%% SECTION 2 %%%%%%%%%
%%%%%%%%%%%%%%%%%%%%%%%%%%%%

\section{Effective braneworld geometry in Kerr--Schild form}
\label{sec:braneworld-KS}

The gravitational dynamics induced on a 3-brane embedded in a five-dimensional bulk can be obtained by projecting the bulk Einstein equations onto the brane \cite{Shiromizu:1999wj}. Denoting the induced four-dimensional metric by $g_{\mu\nu}$, the effective field equations in the absence of bulk matter fields can be written as
\be
\label{eq:effective-brane-equations-general}
G_{\mu\nu} = - \Lambda_{4}g_{\mu\nu} + \kappa_{4}^{2}T_{\mu\nu} + \kappa_{5}^{4}S_{\mu\nu} - E_{\mu\nu}\, ,
\ee
where $T_{\mu\nu}$ is the energy-momentum tensor of matter localized on the brane and $S_{\mu\nu}$ contains contributions quadratic in $T_{\mu\nu}$. Here, $E_{\mu\nu}$ denotes the electric part of the five-dimensional Weyl tensor, encoding nonlocal gravitational effects associated with the bulk geometry and satisfying $E^\mu{}_\mu=0$.

We shall consider the Randall--Sundrum tuning for which the effective four-dimensional cosmological constant vanishes, and restrict ourselves to a vacuum brane,
\be
\Lambda_{4} = 0\, , \qquad T_{\mu\nu} = 0\, .
\ee
The quadratic term $S_{\mu\nu}$ then vanishes, and the effective equations reduce to
\be
\label{eq:effective-vacuum}
G_{\mu\nu} = -E_{\mu\nu}\, .
\ee
Taking the trace immediately gives
\be
\label{eq:R-zero}
R = 0\, ,
\ee
so that \eqref{eq:effective-vacuum} can equivalently be written as
\be
\label{eq:Ricci-E}
R_{\mu\nu} = -E_{\mu\nu}\, .
\ee

A crucial feature of the effective description is that $E_{\mu\nu}$ is not determined locally by the brane geometry and matter content. In general, the system \eqref{eq:effective-brane-equations-general} is therefore not closed without additional information about the five-dimensional bulk. Nevertheless, the scalar constraint \eqref{eq:R-zero} provides a purely four-dimensional condition on the induced geometry. One may then restrict the class of admissible geometries by imposing additional symmetry and geometric structure and use $R=0$ to determine the corresponding metric function. In the stationary and axisymmetric case considered here, this can be implemented through a Kerr--Schild ansatz, which we shall use below to construct the rotating braneworld black-hole geometry.

\subsection{Kerr--Schild geometry and linearized field equations}
\label{subsec:KS-geometry}

We consider a stationary and axisymmetric induced metric of the Kerr--Schild form \cite{Aliev:2005bi}
\be
\label{eq:Aliev-KS-line-element}
ds^{2} = d\bar{s}^{2} + H(r,\theta)\left(du - a\sin^{2}\theta\,d\varphi\right)^{2}\, ,
\ee
where $y^\mu=(u,r,\theta,\varphi)$ and the background line element is
\be
\label{eq:Aliev-flat-background}
d\bar{s}^{2} = -du^{2} - 2dudr + \Sigma d\theta^{2} +(r^{2} + a^{2})\sin^{2}\theta d\varphi^{2} + 2a\sin^{2}\theta dr d\varphi\, ,
\ee
with
\be
\label{eq:Sigma}
\Sigma = r^{2} + a^{2}\cos^{2}\theta\, .
\ee
Although non-diagonal in these coordinates, $d\bar{s}^{2}$ is simply the Minkowski metric written in oblate spheroidal coordinates. The parameter $a$ characterizes the rotation of the geometry.

In tensor notation, the metric \eqref{eq:Aliev-KS-line-element} can be written as
\be
\label{eq:Aliev-KS-tensor}
g_{\mu\nu} = \bar g_{\mu\nu} + H k_{\mu}k_{\nu}\, ,
\ee
where the Kerr--Schild one-form is
\be
\label{eq:Aliev-k-one-form}
k_{\mu}dx^{\mu} = du - a\sin^{2}\theta\,d\varphi\, .
\ee
Raising the index with the background metric gives
\be
\label{eq:Aliev-k-contravariant}
k^{\mu} = \left(0,-1,0,0\right), \qquad k^{\mu}\partial_{\mu} = -\partial_{r}\, .
\ee
The Kerr--Schild vector is null and geodesic with respect to the background metric,
\be
\label{eq:Aliev-null-geodesic}
k^{\mu}k_{\mu} = 0\, , \qquad k^{\nu}\bar\nabla_{\nu}k^{\mu} = 0\, .
\ee
The latter relation follows directly from $k^{\mu}\partial_{\mu}=-\partial_{r}$ together with $\bar\Gamma^{\mu}{}_{rr}=0$. As usual for a Kerr--Schild geometry, these properties are also preserved with respect to the full metric $g_{\mu\nu}$.

The remaining scalar function $H(r,\theta)$ is constrained by the effective vacuum equations. Substituting the Kerr--Schild ansatz into the scalar constraint \eqref{eq:R-zero} gives
\be
\label{eq:Aliev-H-equation}
\left(\partial_r^{2} + \frac{4r}{\Sigma}\partial_{r} + \frac{2}{\Sigma}\right)H = 0\, ,
\ee
or, equivalently,
\be
\label{eq:Aliev-H-compact}
\partial_{r}^{2}\left(\Sigma H\right) = 0\, .
\ee
A two-parameter solution of this equation describing the rotating braneworld black hole is
\be
\label{eq:Aliev-H-solution}
H(r,\theta) = \frac{2Mr - \beta}{\Sigma}\, ,
\ee
where $M$ and $\beta$ are integration constants. Upon transforming the metric to Boyer--Lindquist coordinates, $M$ is identified from the asymptotic geometry with the mass parameter, while $\beta$ occupies precisely the position of $Q^2$ in the Kerr--Newman solution. In the present case, however, no Maxwell field is present on the brane. Instead, $\beta$ is interpreted as a tidal charge encoding nonlocal gravitational effects associated with the five-dimensional bulk \cite{Aliev:2005bi}.

For a Kerr--Schild metric over a flat background, the exact Ricci tensor can be expressed as
\be
\label{eq:Ricci-KS-full}
R_{\mu\nu} = \left(\delta_\mu{}^\rho + H k_\mu k^\rho\right)R^{(1)}_{\rho\nu}\, ,
\ee
where
\be
\label{eq:Ricci-KS-linear}
R^{(1)}_{\mu\nu} = \frac{1}{2}\bar\nabla_{\lambda}\left[\bar\nabla_{\mu}\left(H k_{\nu}k^{\lambda}\right) + \bar\nabla_{\nu}\left(
H k_{\mu}k^{\lambda}\right) - \bar\nabla^{\lambda}\left(H k_{\mu}k_{\nu}\right)\right]\, .
\ee
The inverse metric is likewise exact
\be
\label{eq:KS-full-inverse}
g^{\mu\nu} = \bar g^{\mu\nu} - H k^{\mu}k^{\nu}\, ,
\ee
and, using the null condition, the mixed Ricci tensor becomes exactly linear in $H$
\be
\label{eq:mixed-Ricci-linear}
R^{\mu}{}_{\nu} = R^{(1)\mu}{}_{\nu}\, , \qquad R^{(1)\mu}{}_{\nu}\equiv\bar{g}^{\mu\rho}R^{(1)}_{\rho\nu}\, .
\ee
This linearity is an exact property of the Kerr--Schild parametrization and does not rely on a weak-field approximation \cite{Monteiro:2014cda}.

The effective gravitational equations \eqref{eq:Ricci-E} can therefore be written as
\be
\label{eq:linearized-effective-equation}
R^{(1)\mu}{}_{\nu} = -E^{\mu}{}_{\nu}\, .
\ee
Once the Kerr--Schild scalar $H$ has been specified, this relation determines the projected Weyl tensor associated with the effective brane geometry. The influence of the five-dimensional bulk is thus encoded in an exactly linear Kerr--Schild equation, which provides the natural starting point for the classical double-copy construction developed in the next section.

%%%%%%%%%%%%%%%%%%%%%%%%%%%%
%%%%%%%% SECTION 3 %%%%%%%%%
%%%%%%%%%%%%%%%%%%%%%%%%%%%%

\section{Classical double copy and the tidal charge}
\label{sec:double-copy}

We now turn to the classical double copy of the rotating braneworld geometry introduced in Sec.~\ref{sec:braneworld-KS}. For stationary Kerr--Schild solutions, the relation between the gravitational and gauge-theory equations can be made manifest by contracting the linearized gravitational equations with a Killing vector. This projection reduces the rank-two gravitational equation to a vector equation that can be identified with the Maxwell equation of the single copy. In the present case, the natural choice is the stationary Killing vector
\be
\label{eq:stationary-Killing}
\xi^\mu = (\partial_u)^\mu = (1,0,0,0)\, .
\ee
For the flat background \eqref{eq:Aliev-flat-background}, this vector is covariantly constant. Moreover, the Kerr--Schild scalar and null vector are invariant along the stationary Killing direction and satisfy
\be
\label{eq:KS-stationary-properties}
\mathcal{L}_{\xi}H = 0\, , \qquad \mathcal{L}_{\xi}k_{\mu} = 0\, , \qquad \xi^{\mu}k_{\mu} = 1\, .
\ee
Since the contraction with $\xi^\mu$ removes one factor of the Kerr--Schild null vector, the resulting equation naturally involves the vector field $A_\mu=Hk_\mu$, identified with the single-copy gauge field. Contracting the linearized Ricci tensor with $\xi^\nu$ and using \eqref{eq:mixed-Ricci-linear} together with the properties above gives
\be
\label{eq:Ricci-Killing-contraction}
R^{(1)\mu}{}_{\nu}\xi^{\nu} = \frac{1}{2}\bar\nabla_{\nu} \left[\bar\nabla^{\mu}\left(Hk^{\nu}\right) - \bar\nabla^{\nu}\left(Hk^{\mu}\right)\right] = -\frac{1}{2}\bar\nabla_{\nu}F^{\nu\mu}\, ,
\ee
with field strength $F_{\mu\nu}=2\bar\nabla_{[\mu}A_{\nu]}=\partial_{\mu}A_{\nu}-\partial_{\nu}A_{\mu}$. This is precisely the structure underlying the stationary Kerr--Schild double copy \cite{Monteiro:2014cda}.

Defining the gauge-theory current through
\be
\label{eq:Maxwell-equation}
\bar\nabla_\lambda F^{\lambda\mu} = j^\mu\, ,
\ee
and using the effective gravitational equation \eqref{eq:linearized-effective-equation}, we obtain
\be
\label{eq:current-Weyl-relation}
j^\mu = 2E^\mu{}_\nu\xi^\nu = 2E^\mu{}_u\, .
\ee
Thus, the source of the single-copy gauge field is directly related to the projection of the bulk Weyl contribution along the stationary Killing direction.

For the rotating braneworld solution, the gauge potential is
\be
\label{eq:Aliev-single-copy}
A_\mu = \frac{2Mr-\beta}{\Sigma}\left(1,0,0,-a\sin^2\theta\right)\, ,
\ee
and the independent nonvanishing components of the field strength are
\begin{align}
F_{ru}
&=
\frac{
2\left[
M(a^2\cos^2\theta-r^2)+\beta r
\right]
}{\Sigma^2}\, ,
&
F_{\theta u}
&=
\frac{
a^2(2Mr-\beta)\sin2\theta
}{\Sigma^2}\, ,
\nonumber\\[1mm]
F_{r\varphi}
&=
-\frac{
2a\sin^2\theta
\left[
M(a^2\cos^2\theta-r^2)+\beta r
\right]
}{\Sigma^2}\, ,
&
F_{\theta\varphi}
&=
-\frac{
a(r^2+a^2)(2Mr-\beta)\sin2\theta
}{\Sigma^2}\, .
\label{eq:F-components}
\end{align}

All indices in the single-copy theory are raised with the flat background metric $\bar g_{\mu\nu}$. Since
\be
\label{eq:background-determinant}
\det\bar g_{\mu\nu} = - \Sigma^2\sin^2\theta\, , \qquad \sqrt{-\bar g} = \Sigma\sin\theta\, ,
\ee
and $F^{\mu\nu}$ is antisymmetric, its covariant divergence can be written entirely in terms of the metric determinant. The Maxwell equation therefore takes the form
\be
\label{eq:Maxwell-divergence-coordinate}
\bar\nabla_{\nu}F^{\nu\mu} = \frac{1}{\Sigma\sin\theta}\partial_\nu \left(\Sigma\sin\theta\,F^{\nu\mu}\right) = j^{\mu}\, .
\ee
A direct calculation yields
\be
\label{eq:single-copy-current}
j^\mu = \frac{2\beta}{\Sigma^3}\left(r^2+a^2(1+\sin^2\theta),
\,0,\,0,\,2a\right)\, .
\ee
In particular, the regular current is entirely proportional to the tidal charge. In the regular region $\Sigma\neq0$, the mass-dependent contribution to the single-copy gauge field satisfies the source-free Maxwell equations.\footnote{This statement concerns the regular region of the Kerr--Schild coordinates. Distributional sources may be supported on the singular set $\Sigma=0$ and are not included in the current \eqref{eq:single-copy-current}.}

The zeroth copy follows from a second contraction with the stationary Killing vector. The zeroth-copy scalar is identified with the Kerr--Schild scalar
\be
\label{eq:zeroth-copy-scalar}
\Phi = H = \frac{2Mr-\beta}{\Sigma}\, .
\ee
Using the covariant constancy of $\xi^\mu$, together with $\xi^\mu A_\mu=H$ and the stationarity of the single-copy gauge field, one finds
\be
\xi_\mu F^{\nu\mu} = \bar\nabla^\nu H\, .
\ee
Contracting the Maxwell equation with $\xi_\mu$ therefore gives
\be
\label{eq:zeroth-copy-equation-general}
\bar\square\Phi = \xi_\mu j^\mu = 2\xi_\mu E^\mu{}_\nu\xi^\nu\, ,
\ee
where $\bar\square=\bar g^{\mu\nu}\bar\nabla_\mu\bar\nabla_\nu$. For the stationary and axisymmetric scalar \eqref{eq:zeroth-copy-scalar} on the flat background \eqref{eq:Aliev-flat-background}, one obtains
\be
\label{eq:zeroth-copy-source}
\bar\square\Phi = -\frac{2\beta\left[r^2+a^2(1+\sin^2\theta)\right]}{\Sigma^3}\, .
\ee

These expressions provide a number of immediate checks. The current \eqref{eq:single-copy-current} is conserved, $\bar\nabla_\mu j^\mu=0$, as follows directly from stationarity, axisymmetry, and $j^r=j^\theta=0$. Moreover, for $\beta=0$ the regular sources of both the single and zeroth copies vanish away from their singular support, as expected for the Kerr solution.

\subsection{The tidal-charge contribution}
\label{subsec:tidal-charge}

The role of the tidal charge becomes particularly transparent by separating the Kerr--Schild scalar into its mass and tidal-charge contributions
\be
\label{eq:H-decomposition}
H = H_M + H_\beta\, , \qquad H_M = \frac{2Mr}{\Sigma}\, , \qquad H_\beta = -\frac{\beta}{\Sigma}\, .
\ee
Since the double-copy equations are linear in $H$, the single and zeroth copies inherit the same decomposition. As shown above, in the regular region the mass-dependent contribution is source-free, while the regular gauge and scalar sources are entirely determined by $H_\beta$. The tidal charge therefore leaves a direct imprint on the sources of the double-copy fields.

The central comparison is with the Kerr--Newman black hole. In Kerr--Schild form, its scalar function is
\be
\label{eq:H-KN}
H_{\rm KN} = \frac{2Mr-Q^2}{\Sigma}\, ,
\ee
with the same flat background and Kerr--Schild null vector as the rotating braneworld solution. At the level of the Kerr--Schild geometry, the two solutions are therefore related by the formal replacement
\be
\label{eq:Q2-beta-replacement}
Q^2\longleftrightarrow\beta\, .
\ee
Consequently, the same replacement relates their single-copy gauge fields, currents, zeroth-copy scalars, and scalar sources. This correspondence reflects the analogous relation on the gravitational side, where the effective Weyl contribution plays the role occupied by the electromagnetic energy-momentum tensor in Kerr--Newman under $Q^2\leftrightarrow\beta$ \cite{Aliev:2005bi,Bah:2019sda,Easson:2022zoh}.

Despite this formal equivalence, the physical interpretation is different. The Maxwell field sourcing the Kerr--Newman geometry is proportional to the electric charge $Q$, whereas the charge-dependent contribution to its Kerr--Schild single copy is proportional to $Q^2$; the two gauge fields should therefore not be identified \cite{Easson:2022zoh}. For the braneworld geometry there is no physical Maxwell field associated with $\beta$ on the vacuum brane. Instead, the tidal charge parametrizes the nonlocal gravitational influence of the five-dimensional bulk through $E_{\mu\nu}$. Moreover, unlike $Q^2$, $\beta$ is not constrained to be positive \cite{Dadhich:2000am,Aliev:2005bi}. In particular, the sector $\beta<0$ has no counterpart under an identification with the square of a real Kerr--Newman electric charge, while both the gauge current \eqref{eq:single-copy-current} and the scalar source \eqref{eq:zeroth-copy-source} reverse sign with $\beta$.

The double copy thus retains the effective imprint of the extra dimension through the tidal charge, including its magnitude and sign, but does not by itself encode the higher-dimensional origin of this parameter. Indeed, the same local Kerr--Schild fields are obtained for $\beta>0$ from Kerr--Newman under $Q^2=\beta$, despite the different physical origins of the corresponding gravitational sources. The construction should therefore be understood as the classical double copy of the induced brane geometry: recovering the bulk geometry responsible for the tidal charge requires information beyond the four-dimensional Kerr--Schild fields.

\subsection{Adding electromagnetic charge on the brane}
\label{subsec:em-charge}

The comparison with Kerr--Newman raises the natural question of what happens when an actual electromagnetic field is present on the brane in addition to the tidal contribution from the bulk \cite{Aliev:2005bi}. In this case, the brane is no longer in vacuum and the general effective equations \eqref{eq:effective-brane-equations-general} become
\be
\label{eq:charged-effective}
G_{\mu\nu} = \kappa_4^2 T_{\mu\nu}^{\rm EM} + \kappa_5^4 S_{\mu\nu} - E_{\mu\nu}\, ,
\ee
where we continue to assume a vanishing effective four-dimensional cosmological constant. The tensor $S_{\mu\nu}$ contains the corrections quadratic in the brane energy-momentum tensor and is therefore suppressed in the low-energy regime in which the energy density is small compared with the brane tension \cite{Shiromizu:1999wj,Maartens:2010ar}.

Neglecting these quadratic corrections, the effective equations reduce to
\be
\label{eq:charged-low-energy}
G_{\mu\nu} = \kappa_4^2 T_{\mu\nu}^{\rm EM} - E_{\mu\nu}\, .
\ee
Since both $T_{\mu\nu}^{\rm EM}$ and $E_{\mu\nu}$ are traceless, the scalar constraint remains $R=0$ at this order. The Kerr--Schild scalar reduces to
\be
\label{eq:H-charged-brane}
H = \frac{2Mr-Q^2-\beta}{\Sigma}\, ,
\ee
which combines the Kerr--Newman contribution associated with an electromagnetic charge $Q$ with the tidal contribution parametrized by $\beta$. At the level of the induced Kerr--Schild geometry, the two contributions therefore enter through the combination
\be
Q_{\rm eff}^2 = Q^2+\beta\, .
\label{eq:Qeff}
\ee

The exact linearity of the Kerr--Schild construction makes the corresponding double copy immediate: the single and zeroth copies and their regular sources follow from the results above under the replacement $\beta\rightarrow Q_{\rm eff}^2$. Thus, at this order, the electromagnetic and tidal contributions enter the induced geometry and its double copies only through the combination $Q^2+\beta$, despite their different physical origins. In particular, the physical Maxwell potential on the brane is proportional to $Q$, whereas the charge-dependent contribution to the single-copy gauge field is proportional to $Q^2$, and the two should not be identified.

An especially simple case is $\beta=-Q^2$. Within the low-energy description above, the electromagnetic and tidal contributions then cancel in the induced Kerr--Schild geometry,
\be
H\big|_{\beta=-Q^2}
=
\frac{2Mr}{\Sigma}\, ,
\ee
so that the metric reduces to the Kerr form and the regular sources of the single and zeroth copies vanish. Nevertheless, the physical Maxwell field on the brane remains nonvanishing for $Q\neq0$. The double copy of the induced geometry is therefore sensitive only to the net gravitational imprint of the electromagnetic and bulk contributions, and does not separately retain their physical origin.

This degeneracy does not persist straightforwardly once the full braneworld corrections are retained. For a general brane energy-momentum tensor, the quadratic contribution in \eqref{eq:charged-effective} is
\be
S_{\mu\nu}
=
-\frac14 T_{\mu\alpha}T_\nu{}^\alpha
+\frac{1}{12}T T_{\mu\nu}
+\frac18 g_{\mu\nu}T_{\alpha\beta}T^{\alpha\beta}
-\frac{1}{24}g_{\mu\nu}T^2\, .
\label{eq:Smunu}
\ee
For the electromagnetic energy-momentum tensor, $T=0$, and its trace therefore reduces to
\be
\label{eq:Strace}
S^\mu{}_\mu = \frac14 T_{\alpha\beta}^{\rm EM}T_{\rm EM}^{\alpha\beta}\, .
\ee
Taking the trace of \eqref{eq:charged-effective} then gives
\be
\label{eq:R-charged-brane}
R = - \frac{\kappa_5^4}{4}T_{\alpha\beta}^{\rm EM}T_{\rm EM}^{\alpha\beta}\, .
\ee
Thus, beyond the low-energy approximation, the scalar constraint $R=0$ used in Sec.~\ref{sec:braneworld-KS} is generically modified by terms quadratic in the electromagnetic energy-momentum tensor. Since $T_{\mu\nu}^{\rm EM}$ is itself quadratic in the Maxwell field, these corrections start at order $Q^4$ \cite{Aliev:2005bi}. While the Kerr--Schild structure is preserved, it would be interesting to explore whether these corrections admit a corresponding double-copy interpretation. We leave this question for future work.

%%%%%%%%%%%%%%%%%%%%%%%%%%%%
%%%%%%%% SECTION 4 %%%%%%%%%
%%%%%%%%%%%%%%%%%%%%%%%%%%%%

\section{Conclusions}
\label{sec:conclusions}

In this work, we have studied the classical double copy of the rotating black-hole geometry induced on the brane in the Randall--Sundrum II scenario \cite{Aliev:2005bi}. For this solution, the induced metric admits a Kerr--Schild representation over a flat background, with scalar function $H=H_M+H_\beta$, where $\beta$ is the tidal charge associated with the nonlocal influence of the bulk, encoded in the projected Weyl tensor $E_{\mu\nu}$. The exact linearity of the mixed Ricci tensor in the Kerr--Schild scalar allows the effective gravitational equations to be mapped directly to the single- and zeroth-copy equations.

We found that the single-copy gauge field satisfies a Maxwell equation with a source entirely proportional to the tidal charge, while the mass-dependent contribution is source-free in the regular region. The zeroth-copy scalar exhibits the same structure, with a source likewise proportional to $\beta$. Thus, the bulk contribution encoded by the tidal charge appears directly in the regular sources of both copies. This result becomes particularly transparent by comparison with Kerr--Newman \cite{Bah:2019sda,Easson:2022zoh}. At the level of the Kerr--Schild geometry and its double copies, the two solutions are related by the formal correspondence $Q^2\leftrightarrow\beta$. Their physical origins, however, are different: $Q$ is an electromagnetic charge, whereas $\beta$ encodes the influence of the five-dimensional gravitational field on the brane. The double-copy fields retain the magnitude and sign of the tidal contribution but do not determine its higher-dimensional origin. This distinction is especially relevant for $\beta<0$, which has no counterpart in terms of a real Kerr--Newman electric charge \cite{Dadhich:2000am,Aliev:2005bi}.

We have also considered the effect of adding an electromagnetic charge to the brane. In the low-energy regime, where terms quadratic in the brane energy-momentum tensor can be neglected, the electromagnetic and tidal contributions enter the induced Kerr--Schild geometry through the combination $Q_{\rm eff}^2=Q^2+\beta$, which also determines the single and zeroth copies and their regular sources. Thus, the induced Kerr--Schild double copy cannot distinguish between the local electromagnetic field on the brane and the nonlocal gravitational influence of the bulk. An interesting situation arises for $\beta=-Q^2$, where the two contributions cancel at this order, reducing the induced geometry to Kerr and eliminating the regular double-copy sources despite the presence of a nonvanishing physical Maxwell field. In this case, the cancellation shows explicitly that the double copy of the induced geometry is sensitive only to the net gravitational effect on the brane, even though the underlying electromagnetic and bulk contributions remain separately nonvanishing. Beyond the low-energy approximation, the quadratic contribution $S_{\mu\nu}$ has a nonvanishing trace for electromagnetic matter and modifies the condition $R=0$. The simple $Q^2+\beta$ degeneracy therefore need not persist when the full braneworld corrections are included.

The present construction concerns the induced four-dimensional geometry and does not require knowledge of its complete five-dimensional bulk completion. A natural extension is to study the classical double copy of the full charged braneworld solution, including the contribution generated by the quadratic term $S_{\mu\nu}$, and determine how these high-energy corrections are encoded in the corresponding single- and zeroth-copy sources. Similar constructions for other braneworld geometries involving additional bulk or brane fields could provide further examples in which local matter and higher-dimensional effects contribute simultaneously to the induced geometry \cite{Nakas:2020sey,Nakas:2020crd,Nakas:2021srr,Neves:2021dqx,Estrada:2024lhk,Estrada:2025ice}. More generally, it would be interesting to understand whether the loss of information exhibited here through the combination $Q^2+\beta$ is specific to this solution or reflects a broader feature of the classical double copy of effective geometries, in which physically distinct sources can produce the same gravitational configuration.

\section*{Acknowledgements}

The authors thank Eric Lescano for useful comments and suggestions on the manuscript. This work was supported by UADE through the Academic Research Project ``Black Hole Physics in Modern Theories of Gravity'' (A26T65). J.A.R. also acknowledges support from CONICET.

\bibliographystyle{JHEP} 
\bibliography{references} 

\end{document}